\documentclass[prb,aps,11pt]{revtex4-1} 

\usepackage{amsmath}  
\usepackage{amsfonts} 
\usepackage{graphicx} 
\usepackage[utf8]{inputenc}
\usepackage{gensymb}
\begin{document}


\title{Acceleration Measurements Using Smartphone Sensors: \\ Dealing with the Equivalence Principle}

\author{Martín Monteiro}
\email{monteiro@ort.edu.uy} 
\affiliation{Universidad ORT, Uruguay}

\author{Cecilia Cabeza}
\email{cecilia@fisica.edu.uy}

\affiliation{Instituto de F\'{i}sica, Facultad de Ciencias, Universidad de la Rep\'{u}blica,
             Igu\'{a} 4225, Montevideo, 11200, Uruguay}

\author{Arturo C. Martí}
\email{marti@fisica.edu.uy}

\affiliation{Instituto de F\'{i}sica, Facultad de Ciencias, Universidad de la Rep\'{u}blica,
             Igu\'{a} 4225, Montevideo, 11200, Uruguay}


\date{\today}

\begin{abstract}
Acceleration sensors built into smartphones, i-pads or tablets can
conveniently be used in the Physics laboratory.  By virtue of
the equivalence principle, a sensor fixed in a non-inertial reference
frame cannot discern between a gravitational field and an accelerated
system. Accordingly, acceleration values read by these sensors must
 be corrected for the gravitational component. A physical
pendulum was studied by way of example, and  absolute acceleration and rotation angle values
were derived from the measurements made by the accelerometer and gyroscope.  
Results were  corroborated by comparison with those obtained by video analysis. The
limitations of different smartphone sensors are discussed.
\end{abstract}

\maketitle 

\section{Statement of the problem}

The use of smartphones and similar devices has spread pervasively
worldwide over the past years. The scope of the smartphone utility has
exceeded that initially envisioned.  The smartphone revolution has
impacted even teaching practices, as various experiments can be readily
carried out  using sensors customarily available in smartphones.  Several recent works  have proposed the use of
smartphones in the conduction of laboratory experiments on mechanics \cite{vogt2012analyzing,Chevrier2013,streepey2013using},
electromagnetism \cite{Silva2012,Forinash2012}, optics \cite{Thoms2013}, oscillations \cite{castro2013using,sans2013oscillations} 
and waves \cite{Orsola2013,Kuhn2013acoustic}.

Some such experiments dealt with mechanics problems; specifically, the
measurement of gravitation  \cite{,kuhn2013smartphones}, the determination of elastic energy and
the study  of simple \cite{vogt2012simple}, physical, or spring \cite{kuhn2012analyzing}  pendulums have been addressed. 
A recent study \cite{Shakur2013} focused on the conservation of the angular momentum using a smartphone equipped with an angular rate
sensor, or gyroscope, mounted on a rotating table. The gyroscope sensor
has also been used for the calculation of rotational kinetic energy in
a physical pendulum \cite{Monteiro2014rotational}.

Acceleration and rotational sensors can be used simultaneously. In one study \cite{Monteiro2014angular}, a smartphone
was placed at different distances from the rotation shaft of measurements of centripetal acceleration and angular
velocity of a smartphone placed at different distances from the rotation shaft of a merry-go-round were correlated with the angular radius by means of linear
regressions. Likewise, in a recent study of a physical pendulum, a smartphone affixed to a bicycle wheel was
subject to both rotational as well as low- and high-amplitude oscillating motion (\textit{i.e.}, spinning
in complete circles in one direction, or around a point of
stable equilibrium, respectively) \cite{Monteiro2014exploring}. In this study, the
sensors provided acceleration and angular velocity measurements with
respect to different axes fixed to the smartphone.  For this, a relatively
simple system with one degree of freedom, a generalized coordinate and the conjugate
momentum were determined, enabling the representation of trajectories in the phase
space. This latter, somewhat abstract concept was thus rendered more
tangible.

Little attention has been paid to the fact that acceleration sensors,
when placed in an accelerated system, actually measure an apparent
acceleration. The absolute or real acceleration (\textit{i.e.}, relative to the
reference frame defined by the laboratory) cannot be readily
determined, as it is not possible to discern experimentally between a system
subject to a gravitational field and a non-inertial one by virtue
of the  equivalence principle. In this work, the real
acceleration and the angle of rotation of the smartphone were
determined based on measurements made by the in-built acceleration and
gyroscope sensors. In the experiment, motion in the system occurred in
only one plane, with only one degree of freedom.  The results obtained
from the smartphone were compared with independent determinations made
by the analysis of video recordings. As the use of smartphones in the
laboratory becomes increasingly widespread, the concepts discussed
in this paper can prove useful to both students and instructors.

\section{Experimental set-up: physical pendulum and sensors}

\begin{figure}[h!]
\centering
\includegraphics[width=12cm]{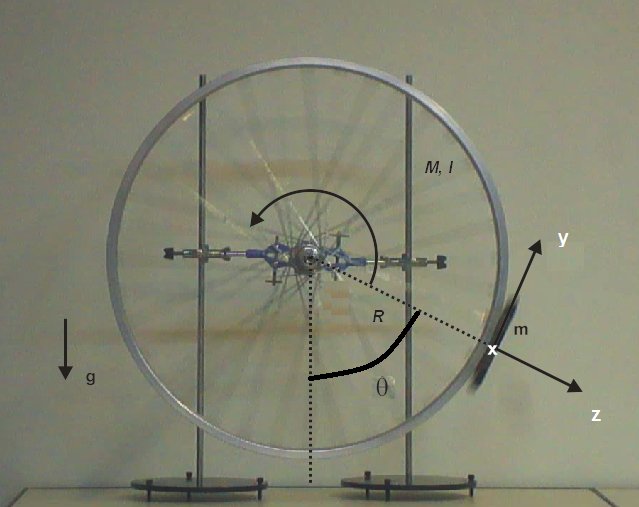}
\caption{Experimental set-up composed of a bicycle wheel with the hub
fixed on a horizontal axis and a smartphone fixed on the outer edge of the tire. The coordinate axes used
in the study are shown.}
\label{fig01}
\end{figure}

A physical pendulum is defined as a rigid body rotating in a plane
around a horizontal axis as a result of the effect of gravity. In this
experiment, the physical pendulum is composed of a bicycle wheel with
its axis fixed in a horizontal position around which the wheel rotates
in a vertical plane, and a smartphone affixed to the outer edge of the
tire, as shown in Fig.~\ref{fig01}. An Android operated smartphone (LG G2 D805) 
furnished with a 3-axis LGE accelerometer sensor (STMicroelectronics, 0.001 m/s2 precision) 
and a 3-axis LGE gyroscope (STMicroelectronics, 0.001 rad/s precision) was used. Technical information 
regarding the exact location of the sensors within the smartphone was obtained from the
manufacturer and verified by physical methods \cite{Monteiro2014exploring}. The Adrosensor
application was used to record sensor readings \cite{playgoogle}.

\begin{figure}[h!]
\centering
\includegraphics[width=8cm]{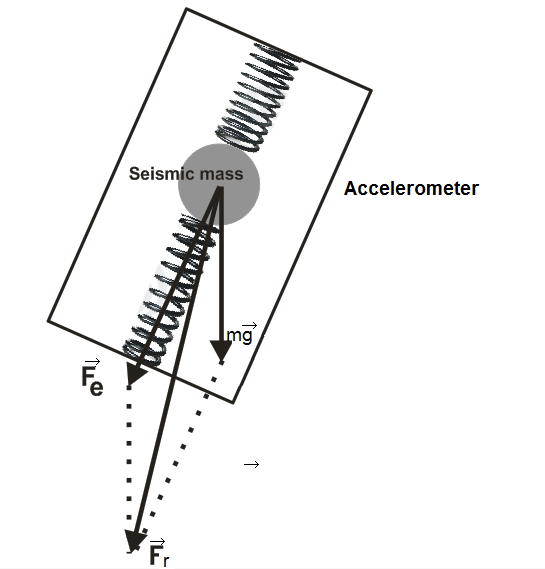}
\caption{Diagram of an acceleration sensor, showing a test particle
and the elastic ($\vec{F}_e$), gravitational ($m \vec{g}$), and resultant ($\vec{F}_R$) forces
acting upon it.}
\label{fig02}
\end{figure}

To make full use of the in-built sensors it is necessary to analyze
their basic operation principles. The construction characteristics of
acceleration sensors are such that they are, actually, force sensors \cite{wikisensor,vogt2012analyzing}. 
These sensors measure the normal force exerted on a test particle (or seismic particle) by a piezoelectric ceramic or
micromechanical capacitor, as shown in Fig. \ref{fig02}. Thus, to
obtain a measurement of the real acceleration of the smartphone it is
necessary to subtract the gravitational component (m\textbf{g}), as shown in Figure \ref{fig02}. This transformation can be readily made if the smartphone is
at rest or in uniform rectilinear motion. In contrast, if the device
is subject to acceleration in an arbitrary direction, 
supplementary measurements are needed by virtue of the equivalence principle.

In addition to an accelerometer, a gyroscope sensor was
used in this experiment. Initially, gyroscopes were based on
rotational gimbal-mounted mechanical devices. Today, smartphones are
equipped with Micro-machined Electro-Mechanical Systems (MEMS) which
measure the Coriolis force on a vibrating body. These sensors provide
direct readings of the angular velocity of the smartphone relative to
predefined axes fixed in the reference frame of the device.

Also linear acceleration and orientation pseudosensors are available
in many smartphone models. Linear acceleration pseudosensors are
supposed to provide readings of the acceleration that the device is
subject to after subtracting the gravitational component. The
orientation pseudosensor integrates the data acquired by several
sensors, including a geomagnetic field sensor, to yield a measurement
of the orientation of the device. In this paper, the results from the accelerometer and the gyroscope are
compared with measurements obtained using these pseudosensors, and the
accuracy of the latter discussed.

The components of vectorial magnitudes are usually read on three axes
($x,y,z$) oriented as if drawn on the smartphone screen. The
measurements used in this study were read by the gyroscope sensor on the $x$ axis and by the acceleration sensor on the $y$ and $z$
axes, for tangential and radial acceleration,
respectively. The recorded data can be downloaded to a computer and
analyzed using suitable software.  An independent measurement of the
system’s motion was available from video data acquired by a digital
camera positioned frontally. The center of the focal field was
positioned at the axle of rotation of the wheel in order to minimize
parallax error. Based on the distance between the axle and the inner
edge of the tire as the length scale, the system’s motion was analyzed
with Tracker software \cite{brown2009tracker}.

\section{ Absolute acceleration and rotation angle}

The time evolution of the rotation angle $\theta$ measured from the point
of stable equilibrium, as shown in Fig.~\ref{fig01}, is derived from Newton’s
Second Law. Neglecting the friction term, the system’s equation of
motion is given by
\begin{equation}
 - m g R \sin \theta = I \ddot{\theta}
 \label{eq1}
\end{equation}
where $m$ is the smartphone mass, $R$ is the distance from the center of
mass, and $I$ is the moment of inertia of the system composed by the
wheel and the smartphone.

The acceleration of the smartphone in the laboratory reference frame is
\begin{equation}
 \vec{a} = - R \dot{\theta}^2 \hat{e}_{r} + R \ddot{\theta} \hat{e}_{\theta}
 \label{eq2}
\end{equation}
where $R$ is the distance from the center of rotation to the center of
mass of the smartphone, located in close proximity to the sensors.
The selected radial and tangential versors, $\hat{e}_{r}$ and $\hat{e}_{\theta}$, coincide with the $z$
and $y$ axes, respectively, on the smartphone. The
gyroscope sensor for the $x$ axis measures directly the angular
velocity on that axis \cite{wikisensor}, so that 
\begin{equation}
\omega_x = - \dot{\theta}
\label{eq3}
\end{equation}
where the  sign is due to the orientation of the axes. It should be
noted in Fig.~\ref{fig01} that the $x$ axis is in the inward direction, while
the sense of rotation is given by the value on the $y$ axis, which in
this case is positive (anticlockwise).

The acceleration value measured by the acceleration sensor, however,
is not a measurement of the real acceleration observed in the
laboratory but of an apparent acceleration, $\vec{a'}$, resulting from the
vectorial sum of the real acceleration and the acceleration associated
with a gravitational field in the opposite direction to that of the
real gravitational acceleration, as follows
\begin{equation}
\vec{a'}=  \vec{a} - \vec{g}.
\label{eq4}
\end{equation}

The components of the apparent acceleration measured by the sensor along
axes $y$ and $z$ of the smartphone are
\begin{equation}
 {a'_y}=R\ddot{\theta}+g\, {\sin \theta},
 \label{eq5}
\end{equation}
\begin{equation}
 {a'_z}=-R\dot{\theta}^2-g\, {\cos \theta}.
 \label{eq6}
\end{equation}
Equations (\ref{eq2}), (\ref{eq3}) and (\ref{eq6}) can be
worked out to yield one of the projected positions as a function of
the smartphone measurements,
\begin{equation}
 \cos \theta=-\frac{{a'_z+R\omega_x}^2}{g},
 \label{eq7}
\end{equation}
while the other projection, derived from equations (\ref{eq1}), (\ref{eq2}) and (\ref{eq5}),
gives 
\begin{equation}
 \sin\theta=-\frac{a'_y}{g(1-\frac{mR^2}{I})}.
 \label{eq8}
\end{equation}
Thus, combining equations (\ref{eq7}) and (\ref{eq8}), the system’s generalized
coordinate can be obtained.

It is worth noting that the denominator of equation (\ref{eq8}) is always
positive, since the moment of inertia of a system (the wheel and
smartphone) is always greater than the moment of inertia of one of its
parts, $I>mR^2$. The limit case where $I=mR^2$ corresponds to a simple
pendulum, and equation (\ref{eq8}), which is only valid for a physical pendulum,  would be indeterminate.

\section{Results}

\begin{figure}[h!]
\centering
\includegraphics[width=14cm]{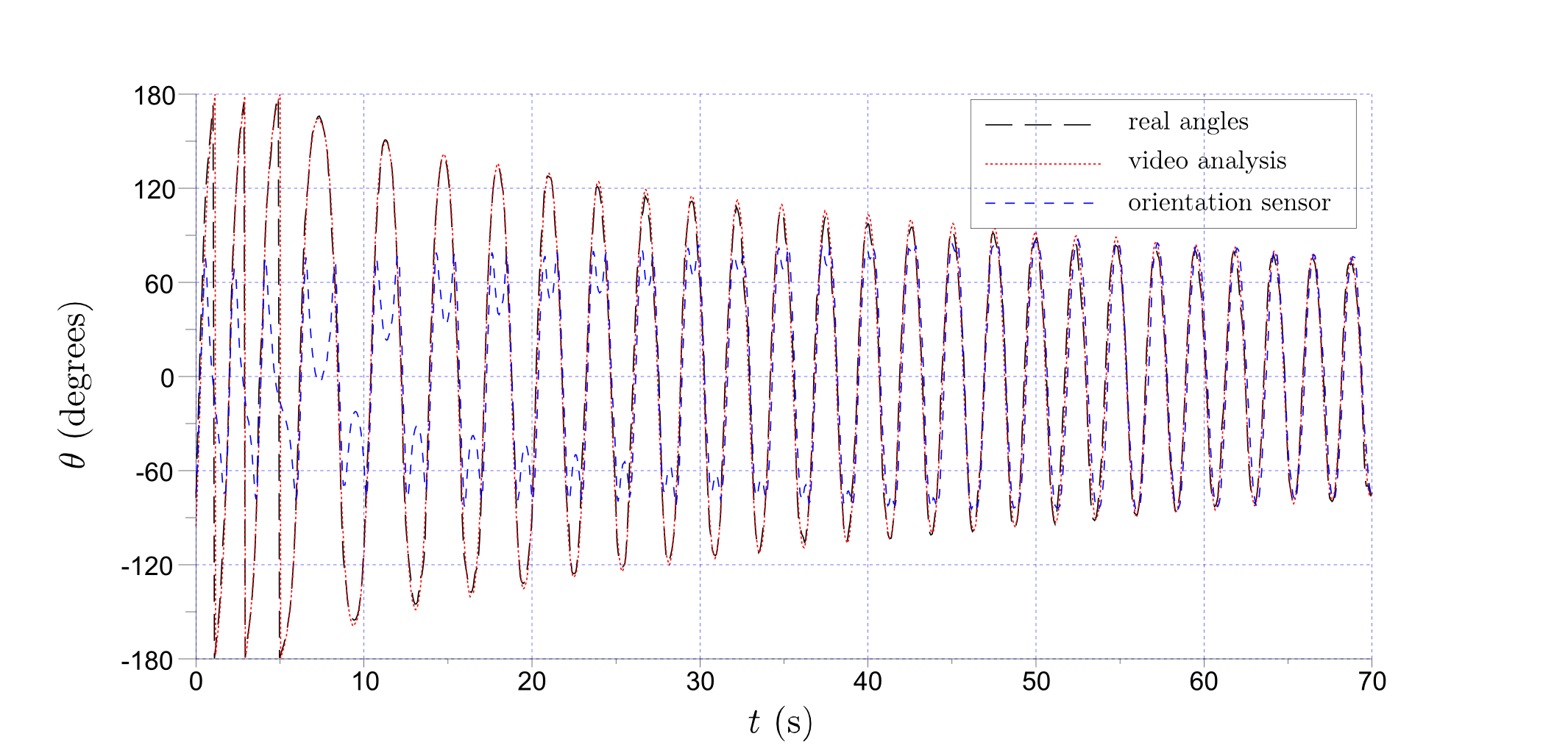}
\caption{
Time evolution of the rotation angle. Angles calculated from
equations (\ref{eq7}) and (\ref{eq8}), derived by Tracker analysis of the video
recording, and measured those by the orientation pseudosensor are shown.}
\label{fig03}
\end{figure}

To analyze the system dynamics, the physical pendulum was set in
motion with sufficient energy to rotate in complete cycles in one
direction. The movement was recorded using the sensors fitted in the
smartphone as well as the video recorder. Figure~\ref{fig03} shows the time
evolution of both the rotation angle calculated by equations (\ref{eq7}) and (\ref{eq8})
as well as that obtained by video analysis using Tracker. A
third measurement read by the orientation pseudosensor is also
shown. The procedure described in the above section yielded results in
agreement with measurements resulting from the analysis of video data
throughout the experiment. The measurements made by the orientation
sensor were in agreement with these results only for angles below 90$ \degree$,
a fact ascribed to the definition of axes in the
orientation pseudosensor.

\begin{figure}[h!]
\centering
\includegraphics[width=14cm]{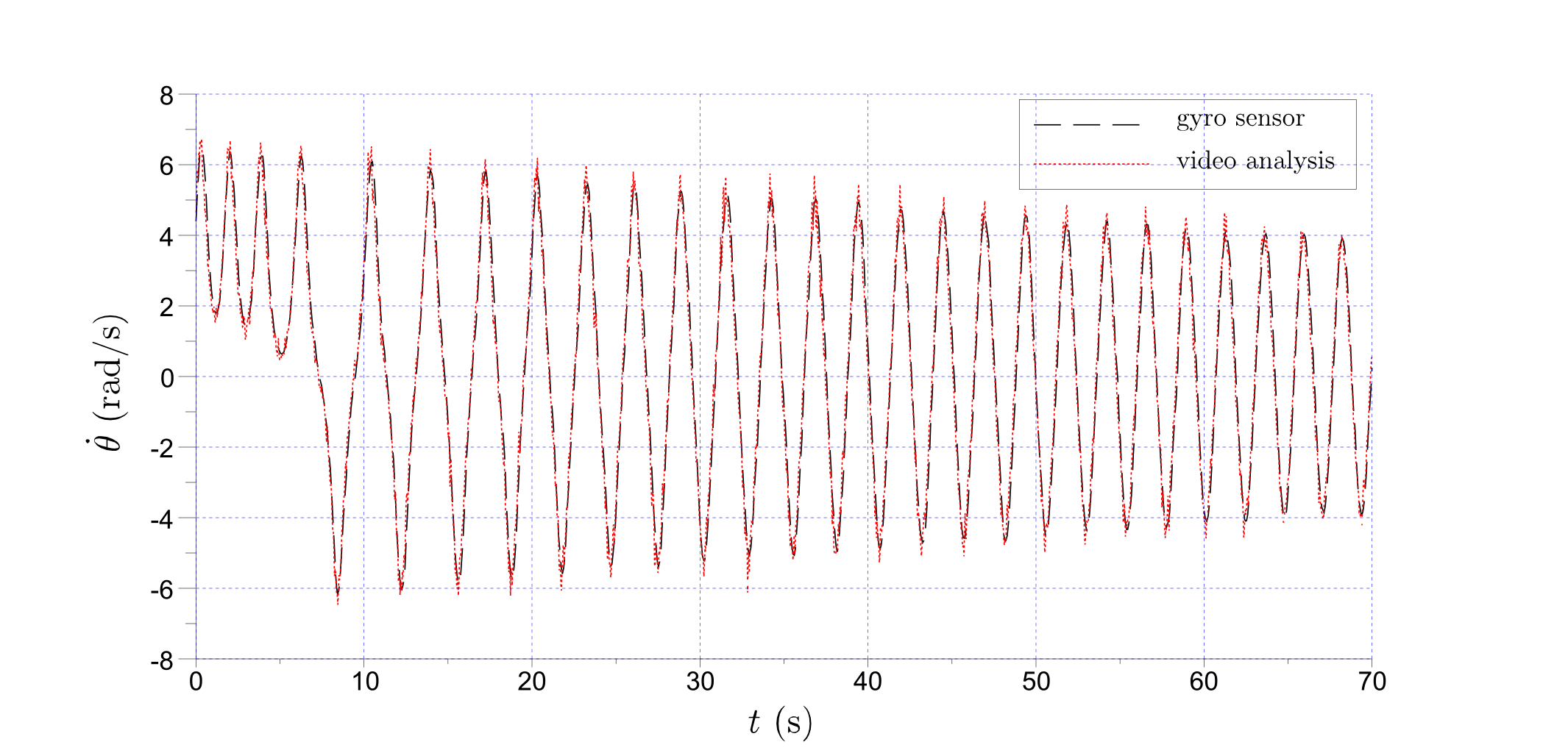}
\includegraphics[width=14cm]{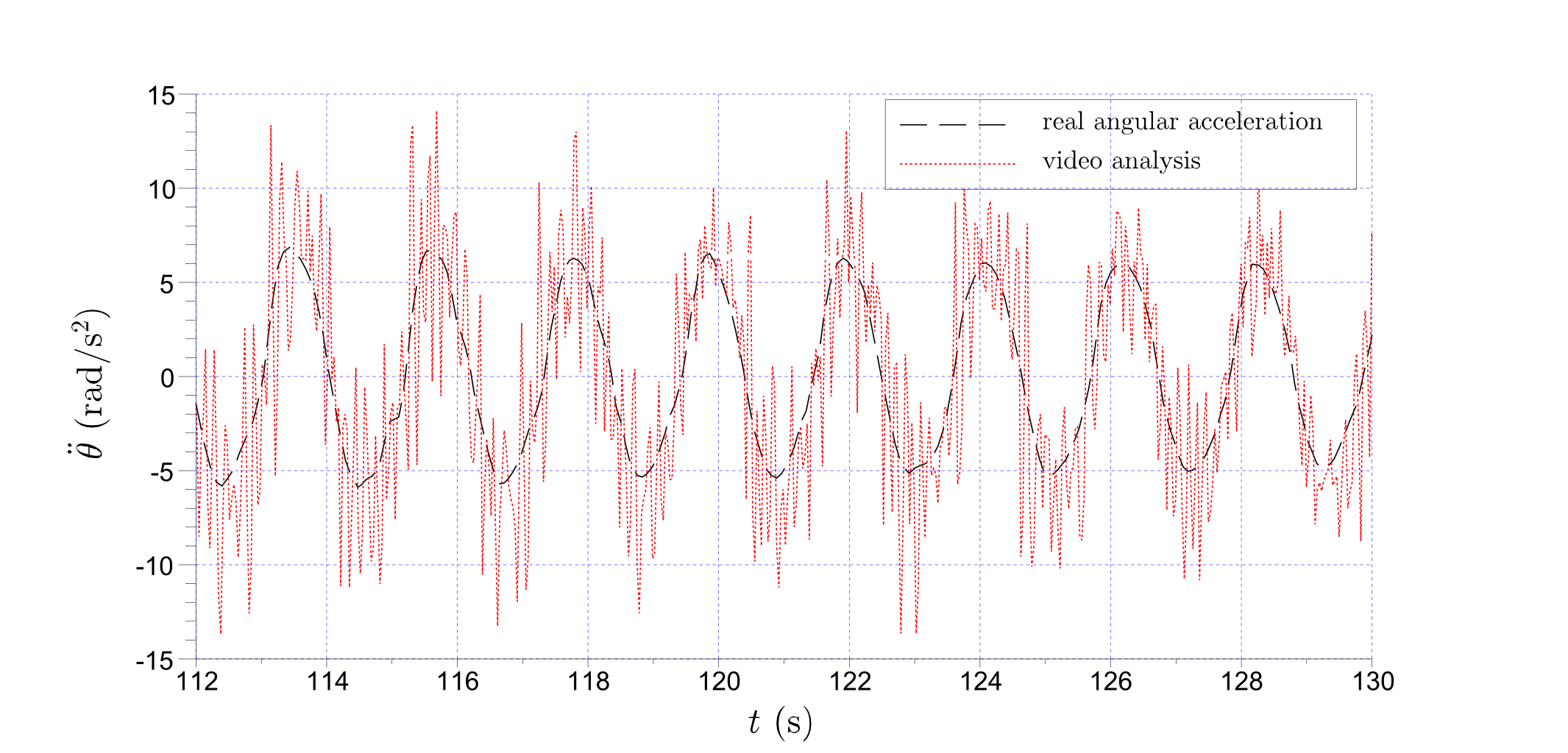}
\caption{ Comparison of gyroscope measurements and results of video
analysis. Time evolution of angular velocity (above) and angular
acceleration (below). Time t=0 was chosen as the first instance at
which the wheel came to a halt.}
\label{fig04}
\end{figure}

Values of angular velocity and acceleration as a function of time read
by the gyroscope sensor and those determined by video
analysis are shown in Fig. \ref{fig04}. Using the gyroscope sensor, angular
velocity is directly read by the sensor whereas the analysis of video
required the numerical calculation of the derivative of the
angle. Angular accelerations shown in the figure (bottom panel) corroborate the
overall agreement between both procedures. However, the numerical
calculation of the derivative, the loss of precision due to the
acquisition time of the digital camera and the task of locating the
object on each image introduce a noise component in the data of
angular velocity and, especially, acceleration, compared with the
measurements made directly by the gyroscope sensor.

\begin{figure}[h!]
\centering
\includegraphics[width=14cm]{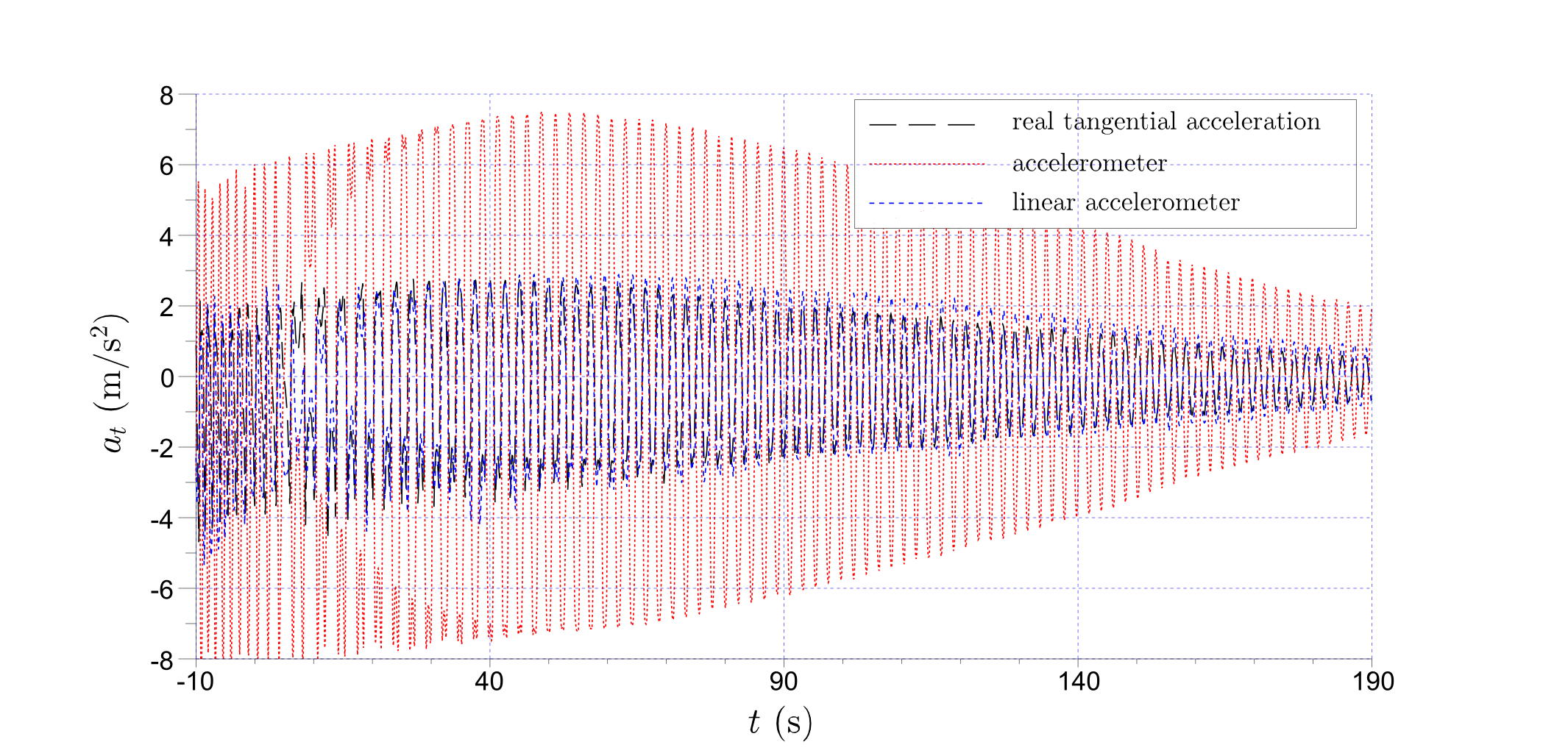}
\includegraphics[width=14cm]{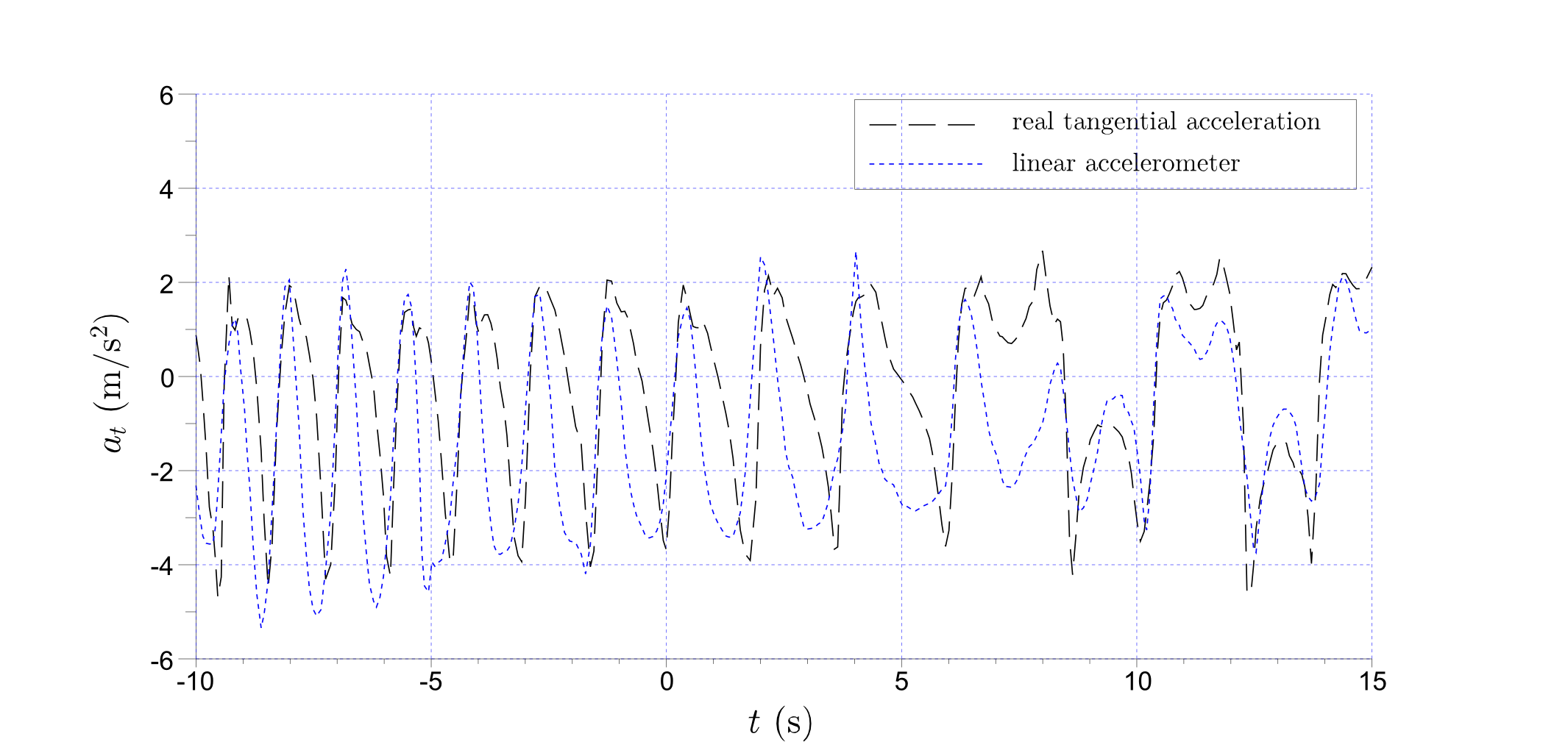}
\includegraphics[width=14cm]{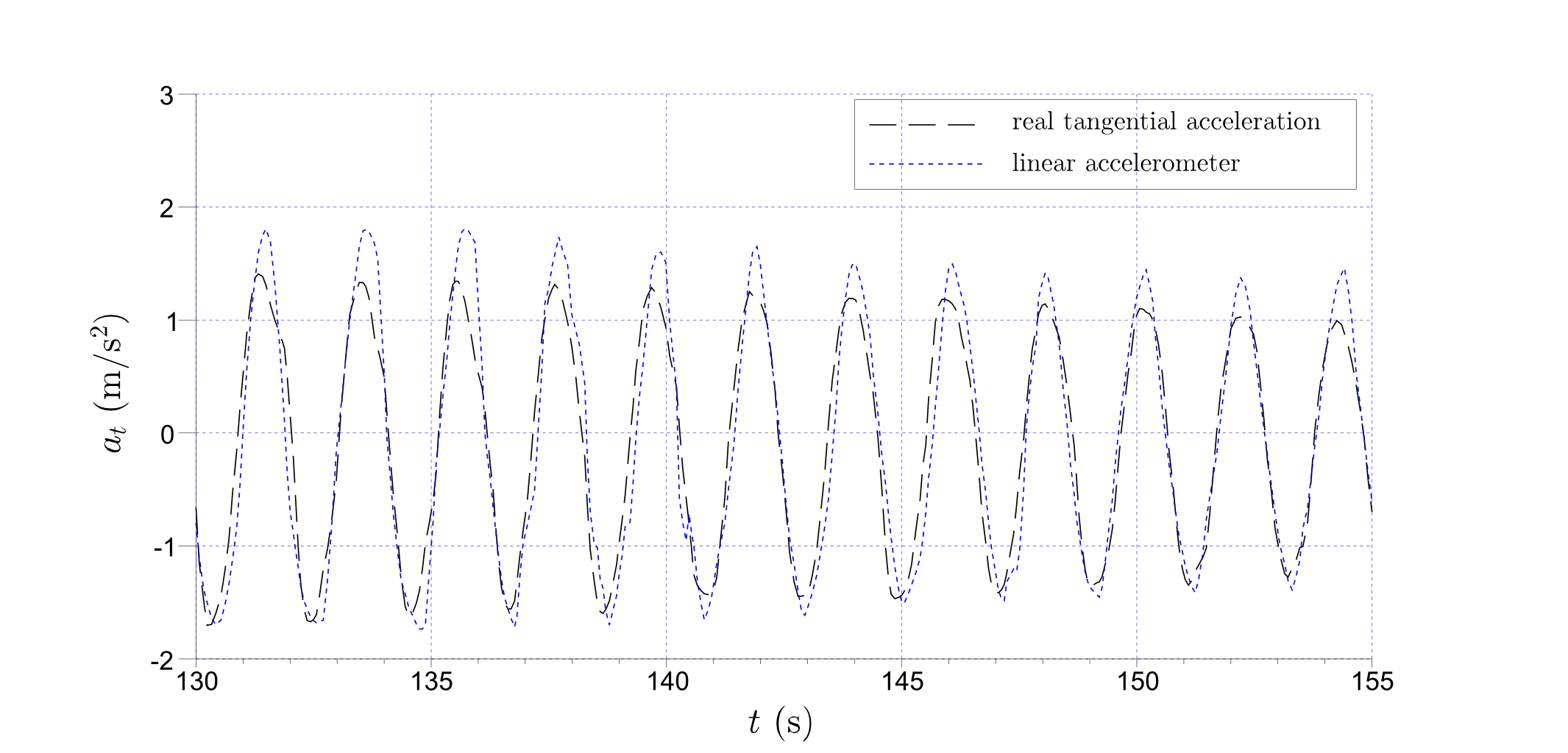}
\caption{ Time evolution of the tangential acceleration value
calculated according to the procedure described above. This value is
compared with those read by the accelerometer and the linear
accelerometer in the smartphone, both of which were inaccurate. The
time evolution throughout the experiment is shown in (a), while (b)
and (c) are enlargements for two different time intervals.}
\label{fig05}
\end{figure}

\begin{figure}[h!]
\centering
\includegraphics[width=14cm]{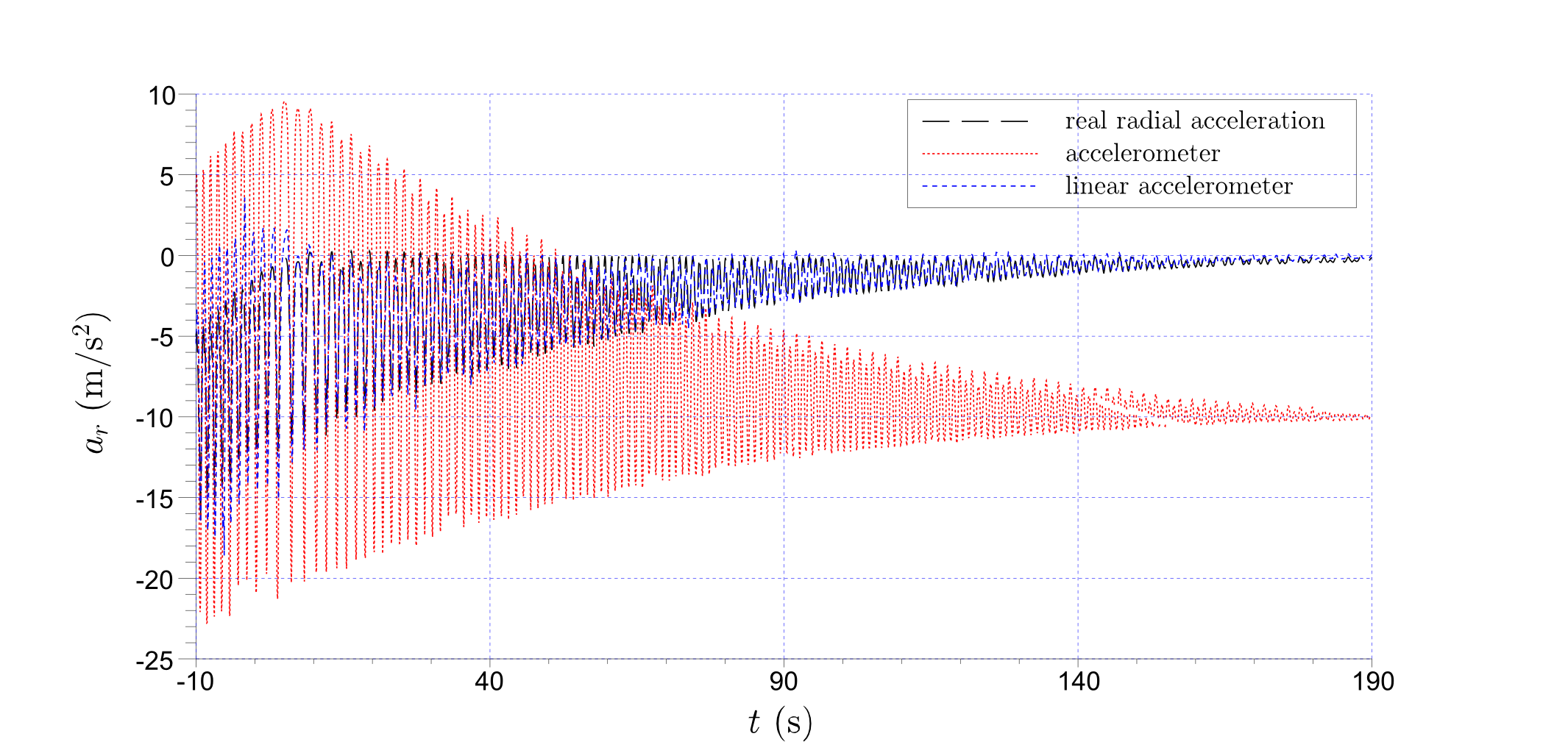}
\includegraphics[width=14cm]{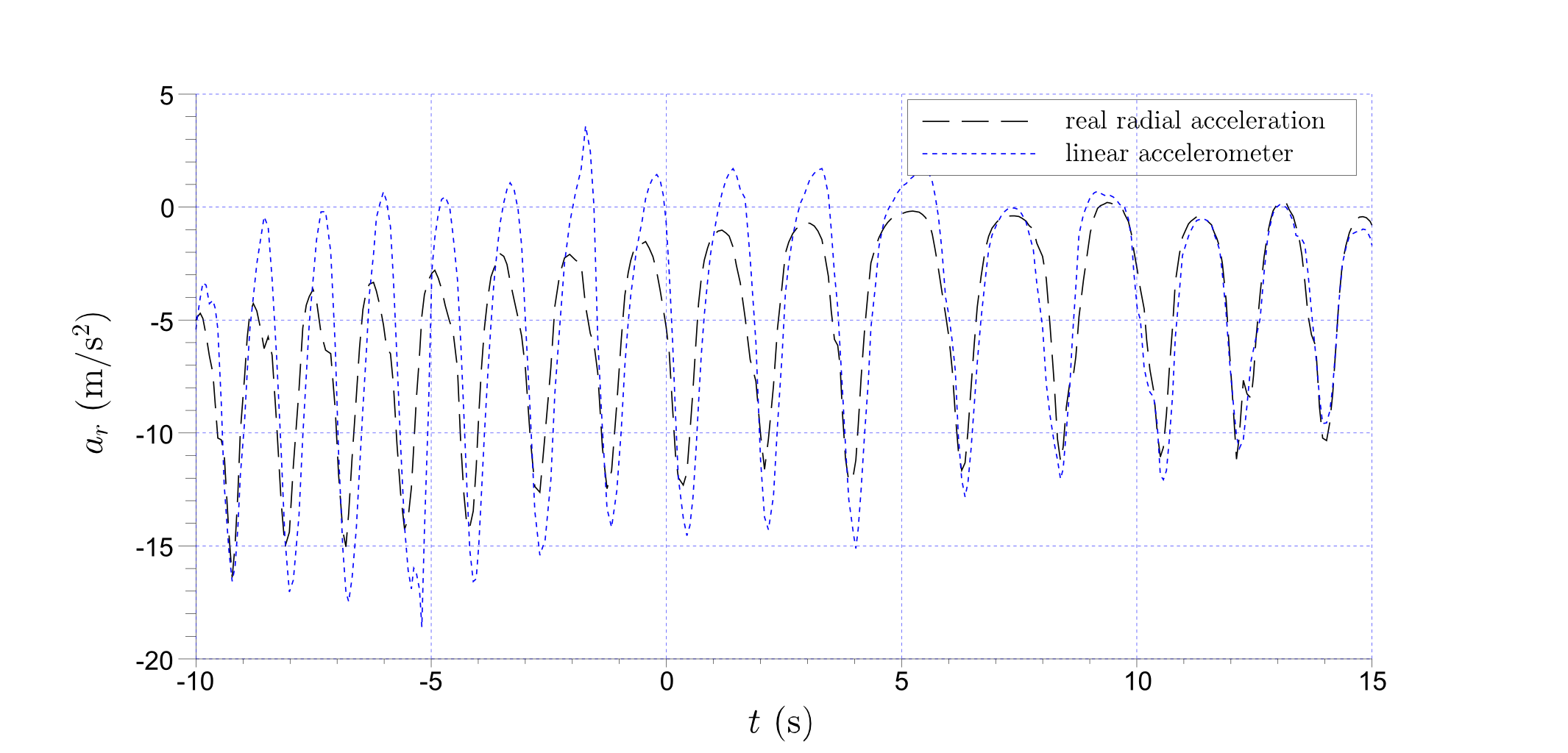}
\includegraphics[width=14cm]{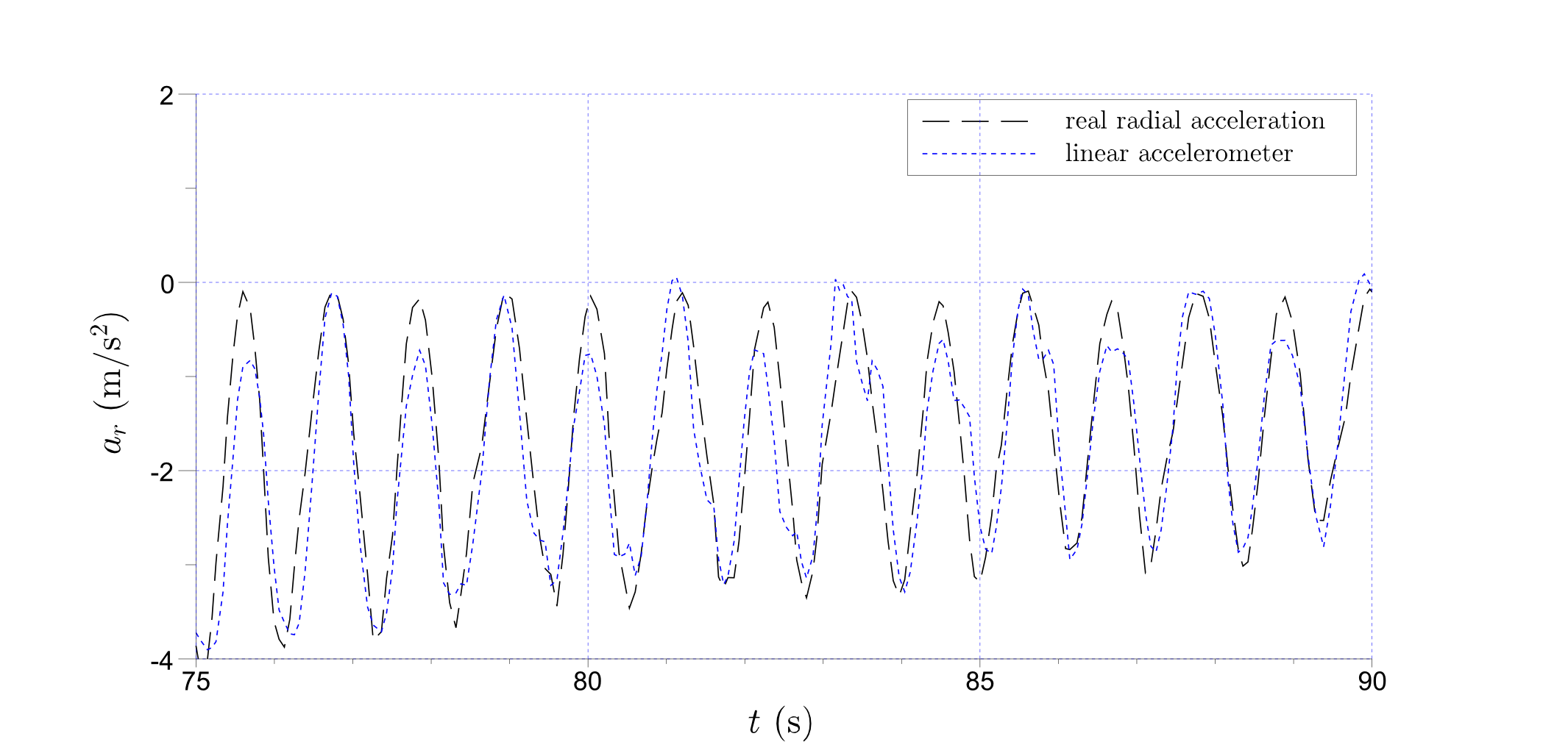}
\caption{ Comparison of radial acceleration, as in Fig. \ref{fig05}.}
\label{fig06}
\end{figure}

The radial and tangential acceleration components derived from the
above equations were compared with the linear acceleration reading from
the pseudosensor and the apparent acceleration from the accelerometer, as shown in Figs. \ref{fig05}
and \ref{fig06}. Figure \ref{fig05} shows the evolution of the tangential acceleration
throughout the experiment. The time interval around t=0, where the
wheel first comes to a halt and begins to oscillate, and an interval
around a later point in time, when the wheel oscillates with
intermediate amplitude, are enlarged for illustration purposes. As expected, the apparent
acceleration differs clearly from the real acceleration calculated
according to the procedure described in the previous
section. Likewise, readings from the linear acceleration pseudosensor
were found to be inaccurate, in particular when the smartphone moves
in proximity to the point of stable equilibrium.

Figure \ref{fig06} shows the radial acceleration as a function of time. As was
the case with the tangential acceleration, the calculated absolute
acceleration was found to differ from that read by the sensors. Panel
(a) shows that the accelerometer reading tends to -10 $m/s^2$ when the
wheel is motionless, whereas both the calculated acceleration value
and that read by the linear accelerometer correctly tend to zero. As
shown in panels (b) and (c), readings from the pseudosensor were
inaccurate.

\section{Conclusions and Prospects}

This paper describes how measurements made using acceleration and
gyroscope sensors fitted in smartphones can be used to obtain the
rotation angle and real acceleration of a physical pendulum. Despite
the constraints resulting from application of the equivalence
principle, 
these measurements can be complemented with those from the gyroscope
sensor to yield real acceleration values. This procedure can be corroborated
by comparison with independent measurements determined by video
analysis.

Diverse measurements can be made using sensors built into smartphones
to elucidate a wide range of physical phenomena. An adequate
understanding of the underlying operation principles can shed
important light on the appropriate use of these applications, a fact
which gains in significance as the use of smartphones becomes more
widespread with the expected decrease in cost.

%

\bibliography{/home/marti/Dropbox/bibtex/mybib}
\bibliographystyle{apsrev4-1}

\end{document}